\documentclass[preprint,aps,showpacs,pre,amsmath]{revtex4}
\usepackage{epsfig}
\usepackage{hyperref}

\begin{document}

\title{Tunnelling through finite graphene superlattices: \\ resonance splitting effect}

\author{ C. Huy Pham$^{1}$ \footnote{cpham@sissa.it} and V. Lien Nguyen$^{2,3}$}
\affiliation{ 
$^1$ SISSA/International School for Advanced Study, Via Bonomea 265, I-34136 Trieste, Italy. \\
$^2$Theoretical and Computational Physics Department, Institute of Physics, VAST,  \\
10 Dao Tan, Ba Dinh Distr.,  Hanoi 10000,  Vietnam. \\
$^3$ Institute for Bio-Medical Physics, 109A Pasteur, $1^{st}$ Distr., Hochiminh City, Vietnam. }

\vspace*{1cm}
\begin{abstract}
An exact expression of the transmission probability through a finite graphene superlattice with an arbitrary number of potential barriers $n$ is derived in two cases of the periodic potential: rectangular electric potential and $\delta$-function magnetic potential. Obtained transmission probabilities show two types of resonance energy: barrier-induced resonance energies unchanged as $n$ varies and well-induced resonance energies undergone the $(n - 1)$-fold splitting as $n$ increases. Supported by numerical calculations for various types of graphene superlattices, these analytical findings are assumed to be in equal applied to all of finite graphene superlattices regardless of potential natures [electric or magnetic] as well as potential barrier shapes.
\end{abstract}

\maketitle
\section{Introduction}
Four decades ago Tsu and Esaki have first demonstrated numerically that for a finite semiconductor superlattice with $n$ potential barriers the transmission probability shows the $(n - 1)$-fold resonance splitting \cite{tsu}. Then, this $(n - 1)$-fold resonance splitting rule was analytically proved for finite semiconductor superlattices with periodic potentials of arbitrary profile \cite{liu1,liu2}. In the limit of large $n$, the resonance energies split gradually as $n$ increases would eventually form the minibands that are responsible for privileged transport properties of semiconductor superlattices such as the Bloch oscillations or the Stark ladders phenomena \cite{esaki}.

The massless Dirac-like behavior of charge carriers in graphene brings about unusual transport properties of not only pristine graphene itself, but certainly also graphene-based nanostructures \cite{neto,sarma}. Therefore, graphene superlattices (GSLs), i.e. graphene under periodic potentials, have been extensively studied in a great number of works \cite{park,brey,barbier,pham,ghosh,masir,snyman,dellan,lequi} for periodic potentials of different natures (electric \cite{park,brey,barbier,pham} or magnetic \cite{ghosh,masir,snyman,dellan,lequi}) and different profiles (Kronig-Penney \cite{park,pham,masir,lequi}, cosine \cite{brey} or square \cite{barbier}). These studies are primarily focused on the behavior of the minibands induced by an infinite periodic potential in the vicinity of the Dirac point and the related transport properties. As for finite GSLs, i.e. graphene-based multi-barrier structures, there are only a few works, where the transmission probability and the conductance are calculated for several values of barrier number $n$  \cite{bai,masir2,barb02}. In particular, calculating the transmission probability for the two types of finite magnetic GSLs (with different potential profiles and $n \leq 5$), Lu et al. noticed that the $(n - 1)$-fold resonance splitting identified in the finite semiconductor superlattices is also applied to the magnetic GSLs examined \cite{lu}.

The purpose of this paper is to show that the $(n - 1)$-fold resonance splitting mentioned is truly applied to all of finite GSLs, electric or magnetic, regardless of potential profiles. To this end, using the transfer matrix approach, we have derived an exact expression of the transmission probability across a finite GSL with an arbitrary number of barriers $n$ in two cases of periodic potentials: rectangular electric potential and $\delta$-function magnetic potential. In both cases, obtained transmission probabilities show two types of resonance energy (RE): $(i)$ the barrier-induced REs that are entirely determined by the single-barrier parameters and $(ii)$ the well-induced REs that could be developed only in the energy ranges corresponding to the minibands in the electronic band of the infinite GSL of the same barrier structure. While the barrier-induced REs are completely insensitive to a change in the barrier number $n$ [$n \geq 1$], the well-induced REs undergo the $(n - 1)$-fold splitting as $n$ increases. These analytical findings are fully supported by numerical calculations performed for finite GSLs with periodic potentials of different natures and shapes and, therefore, they are assumed to be in equal applied to all of finite GSLs regardless of potential natures as well as potential barrier shapes. The most impressive reflection of the resonance behavior of transmission probability, including the $(n - 1)$-fold resonance splitting could be found in the conductance which is numerically demonstrated for two types of electric GSLs with rectangular and triangular potential barriers.

The paper is organized as follows. Sec.II is devoted to a systematic study of the transmission probability across electric GSLs that includes $(i)$ to derive an analytical expression of the transmission probability across a finite electric GSL with arbitrary number of rectangular potential barriers, $(ii)$ to numerically calculate the transmission probability across the finite electric GSLs with different numbers of triangular potential barriers, and $(iii)$ to calculate the conductance of the finite electric GSLs examined. Sec.III shows an analytical expression of the transmission probability across a finite magnetic GSL with arbitrary number of $\delta$-function potential barriers. Results obtained in each section are in detail discussed to identify the resonance spectrum showing the $(n - 1)$-fold resonance splitting. The paper is closed with a brief summary in Sec.IV. 
\section{Electric graphene superlattices}
This section is devoted to the finite/infinite GSLs with periodically electric potentials [electric GSLs - EGSLs]. We first derive an analytical expression of the transmission probability $T_n$ for a finite EGSL with an arbitrary number of rectangular potential barriers, $n$.  Obtained expression shows a full resonance spectrum of $T_n$, including the $(n - 1)$-fold resonance splitting discussed. Then, such the resonance spectrum of $T_n$ is numerically recognized for one more kind of finite EGSLs - the EGSL with triangular potential barriers. The section is closed by showing the conductances which simply reflect the resonance behavior of the transmission probabilities calculated. 
\subsection{Analytical expression for EGSLs with rectangular potential barriers}
We consider a finite one-dimensional EGSL with $n$ rectangular barriers grown along the $x$-direction as schematically illustrated in Fig.1. We will be interested in the case when the low energy properties of charge carriers in the structure can be described by the massless Dirac-like Hamiltonian 
\begin{equation}
\label{eq1}
H_{e} \ = \  v_F \vec{\sigma} \hat{p} + V(x) ,
\end{equation}
where $v_F \approx 10^6 \ m s^{-1}$ is the Fermi velocity of carriers in pristine graphene, $\vec{\sigma} = (\sigma_x , \sigma_y )$ are the Pauli matrices, $\hat{p} = (p_x , p_y )$ is the in-plane momentum, and $V(x)$ describes the periodic potential.

In the simplest case of single rectangular barrier $[n = 1]$, solving the Hamiltonian of eq.(\ref{eq1}) gives straightaway the following expression for the transmission probability \cite{barb03} (see Appendix):
\begin{equation}
\label{eq2}
T_1 \ = \ [ \ 1 + \sin^2 (k_B d_B ) [  k_y U / \hbar v_F k_W k_B  ]^2 \ ]^{-1}, 
\end{equation}
where $U$ is the barrier height, $d_B$ is the barrier width, $k_y$ is the $y$-component of the wave-vector (which is unaffected by the one-dimensional potential $V(x)$), and $k_{B(W)}$ is the $x$-component of the wave-vector inside (outside) the barrier region. Given an incident energy $E$, the wave-numbers $k_{B(W)}$ are defined as
\begin{equation}
\label{eq3}
 k_\lambda \ = \ \sqrt{ [(E - \eta . U) / \hbar v_F ]^2 - k_y^2 } \ ; \ \  \eta = 1 \ {\rm or} \ 0 \ {\rm for} \ \lambda = B \ {\rm or} \ W, \ {\rm respectively}. 
\end{equation} 
The way of $k_y$-dependence of $T_1$ in eq.(\ref{eq2}) expresses a fundamental difference in transmission behavior between graphene and conventional semiconductors. If $k_y = 0$ the transmission probability $T_1$ is always equal to unity, regardless of the barrier height as well as the barrier width. That is the so-called Klein tunnelling - a relativistic effect observed in graphene.

On the other hand, given a non-zero value of $k_y$, the transmission probability $T_1$ of eq.(\ref{eq2}) varies with the incident energy $E$ and reaches the maximum value of unity at the energies which satisfy the equality $\sin (k_B d_B ) = 0$. This equality with $k_B$ defined from eq.(\ref{eq3}) yields the REs of the transmission probability $T_1$ for a single rectangular barrier :
\begin{equation}
\label{eq4}
E^{(\pm )}_l \ = \ U  \pm \hbar v_F \sqrt{ k_y^2 + l^2 \pi^2 / d_B^2 } ; \ \  l - {\rm integers} .
\end{equation}
For example, Fig.1 presents the transmission probability $T_1$ of eq.(\ref{eq2}) for the barrier with $U = 8 \Gamma$ and $d_B = 5 \ nm$ at $k_y = 0.1 \ nm^{-1}$ [$\Gamma \equiv \hbar v_F / 2 d_B$, so if $d_B = 5 \ nm$ then $\Gamma \approx 66 \ meV$]. The arrows indicate the two REs, $E_1^{(-)}$ and $E_1^{(+)}$, determined from eq.(\ref{eq4}).

In the opposite limit of large $n$, an infinite periodic potential produces  minibands in the electronic band structure of GSLs. Using the transfer (T) matrix method, it was shown that the electronic band structure problem of infinite EGSLs with rectangular potential barriers is effectively reduced to solving the following transcendental equation for the Bloch wave-number $k_x$ \cite{pham,barb03}:
\begin{equation}
\label{eq5}
\cos (k_x d) \ = \ f ,
\end{equation}
where  
\begin{equation}
\label{eq6}
f \ = \ \cos (k_W d_W ) \cos (k_B d_B ) +  \frac{(U / \hbar v_F )^2 -  (k_W^2 + k_B^2 )}{ 2 k_W k_B }  \sin (k_W d_W ) \sin (k_B d_B )  ,
\end{equation}
$d_W$ is the well width and $d = d_B + d_W$ is the superlattice period. 

Solutions of eq.(\ref{eq5}) directly give the electronic band structure that consists the minibands separated by the band gaps. Fig.2$(a)$ shows, for example, the cut of the band structure along the $(k_y  = 0.1 \ nm^{-1})$-plane, calculated numerically from eq.(\ref{eq5}) for the EGSL with the same barrier parameters as in Fig.1 and the well width $d_W = d_B$. The solid lines describe the minibands which are separated from each other by the band gaps. So far, no relation is noticed between the superlattice minibands/gaps in Fig.2$(a)$ and the single barrier resonance behavior in Fig.1. Further, once the T-matrix is known one can readily calculate the transmission probability and then the transport characteristics such as the conductance and the shot noise spectrum power \cite{pham}.

For a finite EGSL with an arbitrary number of rectangular barriers, $n$, in the way similar to that realized for finite semiconductor superlattices in Ref.\cite{liu1}, we are able to obtain an exact expression of the transmission probability (see Appendix):
\begin{equation}
\label{eq7}
T_n \ = \ [ \ 1 + Q^2 (k_y U / \hbar v_F k_W k_B )^2 \sin^2 (k_B d_B ) \ ]^{-1} ,
\end{equation}
where 
\begin{equation}
\label{eq8}
Q \ = \ \frac{ f_+^n - f_-^n } {2 \sqrt{ f^2 - 1 }}
\end{equation} 
with 
\begin{equation}
\label{eq9}
f_{\pm} \ = \ f  \pm  \sqrt{ f^2 - 1 } ,
\end{equation}
$f$ defined in eq.(\ref{eq6}), and (the power) $n$ being the number of barriers.

The transmission probability expression of eq.(\ref{eq7}) is valid for any finite $n$, including the case of no barrier, $n = 0$, when $T_n \equiv 1$. Particularly, if $n = 1$, the factor $Q$ equals to unity [see eqs.(\ref{eq8}) and (\ref{eq9})] and eq.(\ref{eq7}) is then reduced to eq.(\ref{eq2}). Note that the two factors $k_y$ and $\sin^2 (k_B d_B )$ are shown at the same place in both $T_1$ of eq.(\ref{eq2}) and $T_n$ of eq.(\ref{eq7}). This  implies that the single barrier transmission properties related to these factors, i.e. the Klein tunnelling effect and the REs of eq.(\ref{eq4}), should be equally reserved for all of finite EGSLs, regardless of the barrier number $n$. It should be however emphasized that while the REs of eq.(\ref{eq4}) are reserved for finite EGSLs with an arbitrary number of barriers, $n$, due to the factor $Q^2$ in eq.(\ref{eq7}) the whole resonance spectrum of a finite EGSL should depend on $n$.

Actually, the factor $Q^2$ in eq.(\ref{eq7}) carries all specific resonance features of the finite EGSLs studied. Regarding the definition of $Q$ in eq.(\ref{eq8}) we consider two cases of the quantity $f$. Note here that for a given EGSL and a given incident angle $\theta$, $k_y = k_W \cos \theta$, this quantity is entirely determined by the incident energy $E$. \\      
{\sl In the case of} $f^2 > 1$, both quantities $f_\pm$ of eq.(\ref{eq9}) are real, and therefore $Q^2$ is always positive. The fact that there is nowhere for $Q$ vanished in the ranges of incident energy, corresponding to the condition of $f^2 > 1$, means that in these energy ranges the REs for finite EGSLs are still associated with only the factor $\sin^2 (k_B d_B )$ and, therefore, they are entirely determined by the same expression of eq.(\ref{eq4}). On the other hand, for $f^2 > 1$ the equation (\ref{eq5}) for infinite EGSLs has no real solution of $k_x$. This implies a presence of band gaps at the corresponding energies in the electronic band of infinite EGSLs. So, we arrive at an important point: in the range of energy, where there is a gap in the electronic band of the infinite EGSL, the REs for all finite EGSLs are the same and determined by eq.(\ref{eq4}), regardless of the barrier number $n$, $n \geq 1$. Due to the fact that, given $k_y$, these REs are determined by only the barrier shape (i.e. $U$ and $d_B$), they will be hereafter called the barrier-induced REs. By comparing Fig.1 and Fig.2$(b)$ we can see that the barrier-induced REs are really unchanged as $n$ varies.  \\  
{\sl In the opposite case of} $f^2 < 1$, the quantities $f_\pm$ of eq.(\ref{eq9}) become complex. To search for the $Q$-behavior in this case, it is convenient to write $f$ in the form $f = \cos \varphi$ with $0 < \varphi < \pi$. Then, from eqs.(\ref{eq8}) and (\ref{eq9}) we have
\begin{equation}
\label{eq10}
Q \ = \ \frac{\sin n\varphi }{ \sin \varphi } \ ; \ \ \ 0  < \varphi < \pi \ . 
\end{equation}
The transmission probability $T_n$ of eq.(\ref{eq7}) reaches the maximum of unity at the energies making Q vanished. Certainly, the $Q$ of eq.(\ref{eq10}) describes well the   cases of $n = 0$ and $n = 1$ discussed above. For $n = 2$ (double-barrier structure) $Q$ is vanished at the single energy, corresponding to $\varphi = \pi / 2$. That is just the RE of $T_2$. Since this RE could be developed only in presence of the well, we will call it the well-induced RE. Increasing the number of barriers/wells, while the barrier-induced REs of eq.(\ref{eq4}) are firmly unchanged, the well-well correlations cause the well-induced REs split. For a given $n$, clearly, there are $(n - 1)$ values of $\varphi$ making $Q$ of eq.(\ref{eq10}) vanished: $\varphi = (m / n) \pi$ with $m = 1, 2, ..., n -1$. Each of these $\varphi$-values determines a value of $f$, and further, a RE. Thus, the well-induced RE developed originally in the double-barrier structure becomes split into $(n - 1)$ sub-REs as the barrier number $n$ increases. This is just the $(n - 1)$-fold resonance splitting claimed in Refs.\cite{tsu,liu1,lu}. Here, it should be also noted that in the considered case of $f^2 < 1$, the equation (\ref{eq5}) has the real solutions which describe minibands in the electronic band of an infinite EGSL. So, we arrive at another important point: the well-induced REs and their $(n - 1)$-fold splitting could be observed only in the energy ranges corresponding to the minibands in the electronic band of the infinite EGSL with the same periodic potential.

Thus, eq.(\ref{eq7}) describes fully the transmission properties of finite EGSLs with rectangular potential barriers. It seems that there are two types of REs (where the transmission becomes perfect): $(i)$ barrier-induced REs that are entirely determined by the single barrier parameters and are the same for all finite EGSLs, regardless of barrier number $n$ and $(ii)$ well-induced REs that could be developed only in the energy ranges corresponding to the minibands in the electronic band of the infinite EGSL and that undergo the $(n - 1)$-fold splitting as $n$ increases. As a demonstration for these statements we show in Fig.2$(b)$ the transmission probability $T_n$ of eq.(\ref{eq7}) plotted as a function of the incident energy $E$ for finite EGSLs with different number of rectangular barriers, $n$. Clearly, $(i)$ the barrier-induced RE ($E_1^{(-)}$ indicated by the arrow) is the same for all finite EGSLs examined and $(ii)$ the well-induced REs developed in the energy ranges corresponding to the minibands in Fig.2$(a)$ undergo the $(n - 1)$-fold splitting as the barrier number $n$ increases [see the peaks in the energy ranges of  $\approx$ (1.5 to 3.5), (4.5 to 6.5), and (7.8 to 9.8$\Gamma $) in Fig.2$(b)$]. Note that the barrier-induced REs may share the place with well-induced REs in the same energy range, depending on $k_y$  [see the energy range of (1.5 to 3.5$\Gamma $) in Fig.2$(b)$]. Such a coexistence of both types of REs might lead to a mistake in observing the $(n - 1)$-fold resonance splitting effect.
\subsection{Numerical demonstrations for EGSLs with triangular potential barriers} 
For periodic potential barriers other than rectangular ones, the $T_n$-expressions similar to eq.(\ref{eq7}) could be also derived in the same way of T-matrix method as realized above (see the magnetic GSL in the next section as an example). Here, we limit ourselves to presenting numerical calculations for one more electric potential barrier model - the one-dimensional triangular barriers illustrated in Fig.3. In this model, for a single lattice unit, $0 \leq x \leq d$, the potential $V(x)$ in the Hamiltonian of eq.(\ref{eq1}) takes the form $V(x) =  (U / d) x$, where $U$  and $d$ are barrier height and superlattice period, respectively. Note that in this potential model a multi-barrier structure (finite EGSL) is characterized by the three parameters: $U$, $d$, and the barrier number $n$.

In general, the transmission probability across any multi-barrier structure of periodic potentials can be numerically calculated in the way of T-matrix as  suggested in Ref.\cite{chau}. We have in this way calculated the transmission probability $T_n$ for finite EGSLs with different number of triangular barriers, $n$. Results shown in Fig.3 are for the barriers of $U = 8 \Gamma$ and $d = 8 \ nm$ at $k_y = 0.1 \ nm^{-1}$ [ $\Gamma \equiv \hbar v_F / d$]. On the one hand, the resonance spectrum of $T_n$ in this figure is rather similar to that in Fig.2$(b)$. The $(n - 1)$-fold splitting of well-induced REs is clearly recognized [see the peaks in the energy ranges of  $\approx$ (1.6 to 3.2), (4.8 to 6.6), and (7.8 to 9.8$\Gamma $)]. These energy ranges are believed to be corresponding to the minibands in the electronic band of the infinite EGSL with the same periodic potential [by checking the band structure similar to Fig.2$(a)$].

On the other hand, there is an important difference between Fig.3 (for triangular barriers) and Fig.2$(b)$ (for rectangular barriers) in relation to the "barrier-induced" REs. At these energies, all of $T_n$ are equal to unity in Fig.2$(b)$ [perfect transmission], while Fig.3 shows $T_1 < 1$ and even $T_n$ decreasing as $n$ increases [enhanced imperfect transmission]. Such an imperfect transmission at the barrier-induced REs observed in Fig.3 is first related to a smoothness of the triangular potential that partly prevents the Klein tunnelling across the barrier. Additionally, this smooth potential effect should be accumulated with increasing barrier number that makes $T_n$ lowering as $n$ increases. Such the $T_n$-behavior at the barrier-induced REs identified in Fig.3 for finite EGSLs with triangular potential barriers should be found in the resonance spectrum of any finite EGSL with smooth potential barriers.

It is worthy to mention that we have also carried out numerical calculations of $T_n (E)$ for finite EGSLs in the potential models other than those considered above. Obtained results are all similar to Fig.3 in supporting the presence of two types of REs as deduced from eq.(\ref{eq7}). The $(n - 1)$-fold resonance splitting is the property of (only) the well-induced REs and should be observed in the resonance spectrum of any finite EGSL, regardless of the potential barrier profile.  
\subsection{Conductance}
 An accurate reflection of the resonance behavior of transmission probability could be found in the conductance. Given the transmission probability $T(E, \theta )$, the conductance at zero temperature can be calculated within the Landauer formalism:
\begin{equation}
\label{eq11}
G  \ = \  G_0  \int_{-\pi / 2}^{\pi / 2} T(E, \theta ) \cos \theta d\theta ,
\end{equation} 
where $G_0 = 4 e^2 E_F W / \hbar^2 v_F$, $E_F$ is Fermi energy, and $W$ is  the sample size along the $y$-direction. Using eq.(\ref{eq11}) the conductance $G$ has been calculated for two types of finite EGSLs with $T_n$ given in Figs.2 [rectangular barriers] and 3 [triangular barriers]. Obtained results are presented in Fig.4.

In both Figs.4$(a)$ and $(b)$ all three curves of different $n$ reach their highest peaks at the energy close to the barrier-induced RE [ $\approx 1.6 \Gamma$ in $(a)$ and $\approx 1.1 \Gamma$ in $(b)$]. For the rectangular barriers in Fig.4$(a)$ all the three peaks at this energy are equal in height, independently of the barrier number $n$. For the triangular barriers in Fig.4$(b)$, however, in consistency with the transmission probabilities in Fig.3 these peaks are lowered as $n$ increases. 

Importantly, beyond the highest peak at the barrier-induced RE the peaks in $G(E)$-curves at higher energies in both Fig.4$(a)$ and $(b)$ reflect well the $(n - 1)$-fold resonance splitting found in the transmission probability (Again, due to a smoothness of triangular barriers this splitting is less clear in Fig.4$(b)$ compared to Fig.4$(a)$ for sharp barriers of rectangular profile).    

\section{Magnetic graphene superlattices}
Now, we consider the GSLs with periodic magnetic potential barriers [magnetic GSLs - MGSLs]. The $(n - 1)$-fold resonance splitting was numerically demonstrated for two types of MGSLs with step and sinusoidal barriers \cite{lu}. Actually, there is a close relation in electronic properties between corresponding EGSLs and MGSLs \cite{tan}. So, certainly, it is possible to derive analytical expressions of the transmission probability also for the finite MGSLs in the same way as that realized above for the EGSLs.

Indeed, for definition, we consider the case of $\delta$-function magnetic barriers as schematically illustrated in Fig.5. The magnetic field is assumed to be uniform in the $y$-direction and staggered as periodic $\delta$-function barriers of alternative signs in the $x$-direction, so for a single lattice unit the field profile has the form
\begin{equation}
\nonumber
\label{eq12}
\vec{B} \ = \ B_0 [ \delta (x + d_B / 2) - \delta (x - d_B / 2) ] \ \hat{z} ,
\end{equation}  
where $B_0$ is the barrier strength and $d_B$ is the barrier width. The corresponding vector potential $\vec{A}$ in the Landau gauge is
\begin{equation}
\nonumber
\label{eq13}
\vec{A}(x) \ = \ B_0 l_B  \Theta (d_B / 2 - | x | ) \ \hat{y} ,
\end{equation}
where $\Theta (x)$ is the Heaviside step function and $l_B = \sqrt{ \hbar c / e B_0 }$ is the magnetic length. Due to a richness of fundamental electronic properties and a simplicity of mathematical treatment the infinite MGSLs with this $\delta$-function barriers have been extensively studied \cite{ghosh,lequi,masir2}. 

In the case of MGSLs with the vector potential $\vec{A}$, instead of $H_e$ of eq.(\ref{eq1}) we have to deal with the Hamiltonian of the form 
\begin{equation}
\nonumber
\label{eq14}
H_{m} \ = \  v_F \vec{\sigma} ( \hat{p} + e \vec{A}) ,
\end{equation}
where $e$ is the elementary charge. It seems that the transmission probability $T_n$ for the finite MGSLs described by this Hamiltonian can be derived in exactly the same way as that realized above for EGSLs. So, it is reasonable to mention only the differences between the two problems. 

The transmission probability for a single $\delta$-function magnetic barrier takes the form
\begin{equation}
\label{eq15}
T_1 \ = \ [ \ 1 + \sin^2 (k_B d_B ) [  e A_0 E / \hbar^2 v_F k_W k_B  ]^2 \ ]^{-1}, 
\end{equation}
where
\begin{equation}
\nonumber
\label{eq16}
 k_\lambda \ = \ \sqrt{ ( E / \hbar v_F )^2 - ( k_y + \eta . e A_0 / \hbar )^2 } \ ; \ \  \eta = 1 \ {\rm or} \ 0 \ {\rm for} \ \lambda = B \ {\rm or} \ W, \ {\rm respectively}. 
\end{equation}
These expressions are respectively in place of eqs.(\ref{eq3}) and (\ref{eq4}) in the case of EGSLs. Note that in difference from eq.(\ref{eq3}) there is no $k_y$-factor in the second term in $T_1$ of eq.(\ref{eq15}). So, the transmission probability through a single $\delta$-function magnetic barrier might be finite even at $k_y = 0$ [The $k_y$-dependence of $T_1$ (\ref{eq15}) is numerically demonstrated in ref.\cite{masir2}].

The transmission probability $T_1$ of eq.(\ref{eq15}) shows the REs
\begin{equation}
\nonumber
\label{eq17}
E^{(\pm )}_l \ = \  \pm \hbar v_F \sqrt{ (k_y + e A_0 / \hbar )^2 + l^2 \pi^2 / d_B^2 } ; \ \  l - {\rm integers} ,
\end{equation}
which are in similarity to REs of eq.(\ref{eq4}) determined entirely by the single barrier parameters [barrier-induced REs].

The transcendental equation of eq.(\ref{eq5}) is in equal applied for the $\delta$-function magnetic barriers, but the quantity $f$ of eq.(\ref{eq6}) is now replaced by 
\begin{equation}
\label{eq18}
f \ = \ \cos (k_W d_W ) \cos (k_B d_B ) -  \frac{(e A_0 / \hbar )^2 +  (k_W^2 + k_B^2 )}{ 2 k_W k_B }  \sin (k_W d_W ) \sin (k_B d_B )  .
\end{equation}
The only difference between the two quantities $f$ in eq.(\ref{eq6}) (for electric rectangular barriers) and eq.(\ref{eq18}) (for $\delta$-function magnetic barriers) is that the product $k_y U$ in eq.(\ref{eq6}) is replaced by $e A_0 E / \hbar$ in eq.(\ref{eq18}). 
 
Further, the transmission probability through a finite MGSL with an arbitrary number of $\delta$-function barriers is obtained in the form
\begin{equation}
\label{eq19}
T_n  \ = \ [ \ 1 + Q^2 (e A_0 E / \hbar^2 v_F k_W k_B )^2 \sin^2 (k_B d_B ) \ ]^{-1} .
\end{equation}
The rest expressions of $Q$ and $f_{\pm}$, eqs.(\ref{eq8}) and (\ref{eq9}), are the same for both EGSL and MGSL problems under study.

It is important to note that while the transmission probabilities $T_n$ of eq.(\ref{eq7}) and eq.(\ref{eq19}) are very different in the $k_y$-dependence, all the factors related to the REs in these two $T_n$-expressions are exactly the same (i.e. $\sin^2 (k_B d_B)$ and $Q^2$). So, all what we have stated about the resonance spectrum of finite EGSLs in the previous section, including the $(n - 1)$-fold splitting of the well-induced REs, are undoubtedly reserved for the finite MGSLs considered.

As a demonstration, we present in Fig.5 the transmission probability $T_n$ of eq.(\ref{eq19}) plotted versus the incident energy $E$ for finite MGSLs with different numbers of $\delta$-function barriers [$A_0 = 0.4$ and $d_B = d_W = 5 \ nm$]. Clearly, like Figs.2 and 3 for finite EGSLs, Fig.5 shows the barrier-induced REs $E_1^{(+)}$ (indicated by the arrow) which are the same ($\approx 0.8 \Gamma$) for all the MGSLs with different $n$. At the same time, obviously, the well-induced REs developed at three different energy ranges undergo the $(n - 1)$-fold splitting.

In order to see the resonance spectrum in a large range of energy all the Figs.2, 3, and 5 are limited to some small values of $n$. As an addition, Fig.6$(a)$ is focused on showing in more detail the $(n - 1)$-fold resonance splitting in a narrow energy range, $\approx 0.5$ to 0.7 $\Gamma$, separated from Fig.5. In this narrow energy range it is possible to distinguish the well-induced resonance peaks even if $n$ is relatively large. Fig.6$(a)$ is a typical demonstration for the $(n - 1)$-fold resonance splitting of interest. Similar pictures could be certainly set in other energy ranges and for various types of MGSLs as well as EGSLs.

Finally, we would like to note that besides the structural parameters $U$, $d_B$, $d_W$, and $n$, the resonance spectrum of a finite GSL also depends on the $k_y$-value. This can be seen, for example, in Fig.6$(b)$, where the transmission probabilities are shown in the same energy range for the same finite MGSL with $A_0 = 0.8, \ d_B = d_W = 5 \ nm$, and $n = 3$, but at different values of $k_y$. This figure obviously demonstrates a strong and unsystematic $k_y$-dependence of the position as well as the half-width of resonance peaks. Due to such the $k_y$-dependence, an appropriate $k_y$-value should be chosen to obtain a clear picture of the $(n - 1)$-fold resonance splitting.
\section{Conclusions}
We have systematically studied the resonance spectrum of the transmission probability through finite EGSLs and MGSLs with different potential barrier shapes. For the finite EGSL with rectangular potential barriers and the finite MGSL with $\delta$-function potential barriers the transmission probability $T_n (E)$ has been derived analytically. Obtained $T_n (E)$-expressions show two types of REs, barrier-induced and well-induced. The barrier-induced REs are entirely determined by the single-barrier parameters [given $k_y$] and remain unchanged as the barrier number $n$ varies [$n \geq 1$]. The well-induced REs developed only in the energy ranges corresponding to the minibands in the electronic band of the infinite GSL of the same type undergo the $(n - 1)$-fold splitting as the barrier number $n$ increases. 

These analytical findings are fully supported by numerical calculations carried out for finite EGSLs/MGSLs with different potential barrier shapes. So, it is reasonable to assume that they should be in equal applied to all of finite GSLs, regardless of potential nature [electric or magnetic] as well as potential barrier shape. Though the $(n - 1)$-fold resonance splitting found in the present work is just that claimed before in Refs.\cite{tsu,liu2,lu}, it is worthy to remark that this splitting is only associated with the well-induced REs. Actually, a typical reflection of the resonance behavior of transmission probabilities $T_n$, including the $(n - 1)$-fold splitting of the well-induced REs, can be found in the conductance.

In fact, the GLSs considered in the present work are the single-layer graphene-based superlattices. We would like also to mention that we have also calculated the transmission probability through the finite bilayer-graphene-based superlattices. Remarkably, calculations performed for two potential models, the electric potential studied in Ref.\cite{huy1} and the magnetic potential studied in Ref.\cite{huy2}, show the resonance spectra with the two types of REs and the $(n - 1)$-fold resonance splitting similar to those presented above for single-layer graphene superlattices. \\
{\bf Acknowledgments} \\
This work was financially supported by Vietnam National Foundation for
Science and Technology Development under Grant No. 103.02-2013.17. \\ \\
\setcounter{equation}{0}
\renewcommand{\theequation}{A.\arabic{equation}}
{\large \bf APPENDIX} \\
Since the transmission probability $T$ can be exactly expressed in terms of ${\cal T}$-matrix elements,
\begin{equation}
\label{A1}
T \ =  \ 1 -  \frac{|{\cal T}_{21}|^2}{|{\cal T}_{22}|^2} ,   \ \ \
\end{equation}
to find $T$ for a structure we should calculate the corresponding ${\cal T}$-matrix. \\
{\sl (1).Transmission probability $T_1$ of eq.(\ref{eq2})}.
In the case of constant potential, $V(x) = V_n = constant$, the wavefunctions of the Hamiltonian of eq. (\ref{eq1}) can be found in the form 
$ \Psi(x,y) = M_n R_n(x) C_n \exp(ik_yy)$ \cite{chau},
where
\begin{equation}
\label{A2}
M_n = 
 \left(   \begin{array}{cc}
1 & 1    \\
\frac{\hbar v_F (k_n + ik_y)}{E - V_n} &  \frac{\hbar v_F (-k_n + ik_y)}{E-V_n}   \end{array}   \right) ,  \ \ \
\end{equation}
\begin{equation}
\label{A3}
R_n = \rm{diag} \left[ e^{ik_nx}, e^{-ik_nx} \right] , \ \ \
\end{equation}
$k_n = \sqrt{[ (E - V_n )/  \hbar v_F ]^2 - k_y^2}$
and 
$C_n = (A_n, B_n)^T$ being the wavefunction amplitude.

So, in solving the Hamiltonian of eq.(\ref{eq1}) for the single rectangular potential defined as 
\begin{equation}
V(x) = \left \{
  \begin{array}{l l}
    U &  \mbox{if $x_0  \le  x  \le  x_0 + d_B$ } , \nonumber \\
    0  &  \mbox{otherwise},  \nonumber
  \end{array} \right.
\end{equation}
the continuity of the wavefunctions at $x=x_0$ and $x=x_0+d_B$ reads:
\begin{eqnarray}
      M_W R_W(x_0) C_1 & = & M_B R_B(x_0) C_2 \nonumber \\
  M_B R_B(x_0 + d_B) C_2 & = & M_W R_W(x_0 + d_B) C_3 . \nonumber 
\end{eqnarray}
Here, $C_1$, $C_2$ and $C_3$ are respectively the amplitudes of wavefunctions in the left, inside, and the right of the barrier;
$M_{W(B)}$ and $R_{W(B)}$ are respectively defined in Eqs. (\ref{A2}) and (\ref{A3}) for $V_n = 0 (U)$.

From the ${\cal T}$-matrix relation, $C_3 = {\cal T} C_1$, the ${\cal T}$-matrix for the single barrier considered can be obtained as
\begin{equation}
\nonumber
{\cal T}(x_0 ) = R_W^{-1}(x_0 + d_B ) M_W^{-1} M_B R_B (d_B ) M_B^{-1} M_W R_W(x_0 ) .
\end{equation}
Regarding the expression of eq.(\ref{A1}), this ${\cal T}$-matrix gives straightaway the transmission probability $T_1$ of eq.(\ref{eq2}). Here, notice that $T_1$ doesn't depend on $x_0$ as it should be. 

The matrix ${\cal T}(x_0)$ has an important property 
\begin{equation}
\label{A4}
{\cal T}(x_0) = R_W^{-1}(x_0){\cal T}(0) R_W(x_0) , \ \ \
\end{equation}
which is useful for calculating the ${\cal T}$-matrix for a multi-barrier structure. \\
{\sl (2).Transmission probability $T_n$ of eq.(\ref{eq7})}. For a finite EGSL with $n$ rectangular barriers the potential $V(x)$ in the Hamiltonian of eq.(\ref{eq1}) has the form
\begin{equation}
V(x) = \left \{
  \begin{array}{l l}
    U & \mbox{if $(j - 1) d  \le  x  \le  (j - 1) d + d_B$},   \nonumber  \\
    0 &  \mbox{otherwise}, \nonumber
  \end{array} \right.
\end{equation}
where $j$ is an integer, $1 \le j \le n$, $U$, $d_B$ and $d$ are defined above.

Actually, the ${\cal T}$-matrix for this multi-barrier potential can be calculated as
\begin{equation}
\label{A5}
{\cal T}_n  \ = \ {\cal T}(nd) ... {\cal T}(2d) {\cal T}(d) {\cal T}(0) . \
\end{equation}
Using eq. (\ref{A4}), we write ${\cal T}_n = R_W^{-n}(d) [R_W(d){\cal T}(0)]^n$, where the matrix $P(d) \equiv R_W(d){\cal T}(0)$ is often called characteristic matrix. It could be shown that \cite{masir2}
\begin{equation}
[P(d)]^n = 
 \left( \begin{array}{cc}
p_{11}Q_n - Q_{n-1} & p_{12} Q_n  , \nonumber  \\
p_{21}Q_n         & p_{22} Q_n - Q_{n-1} \end{array} \right) , \nonumber
\end{equation}
where $Q_n$ is given in eq.(\ref{eq8}) and
$p_{ij}$ ($i,j = 1, 2$) are components of the matrix $P(d)$.

Using the ${\cal T}_n$-matrix of eq.(\ref{A5}), some lengthy, but elementary algebraic calculations give the transmission probability of eq.(\ref{eq7}) .

\newpage
\begin{figure} [t]
\begin{center}
\includegraphics[width=12.0cm,height=9.0cm]{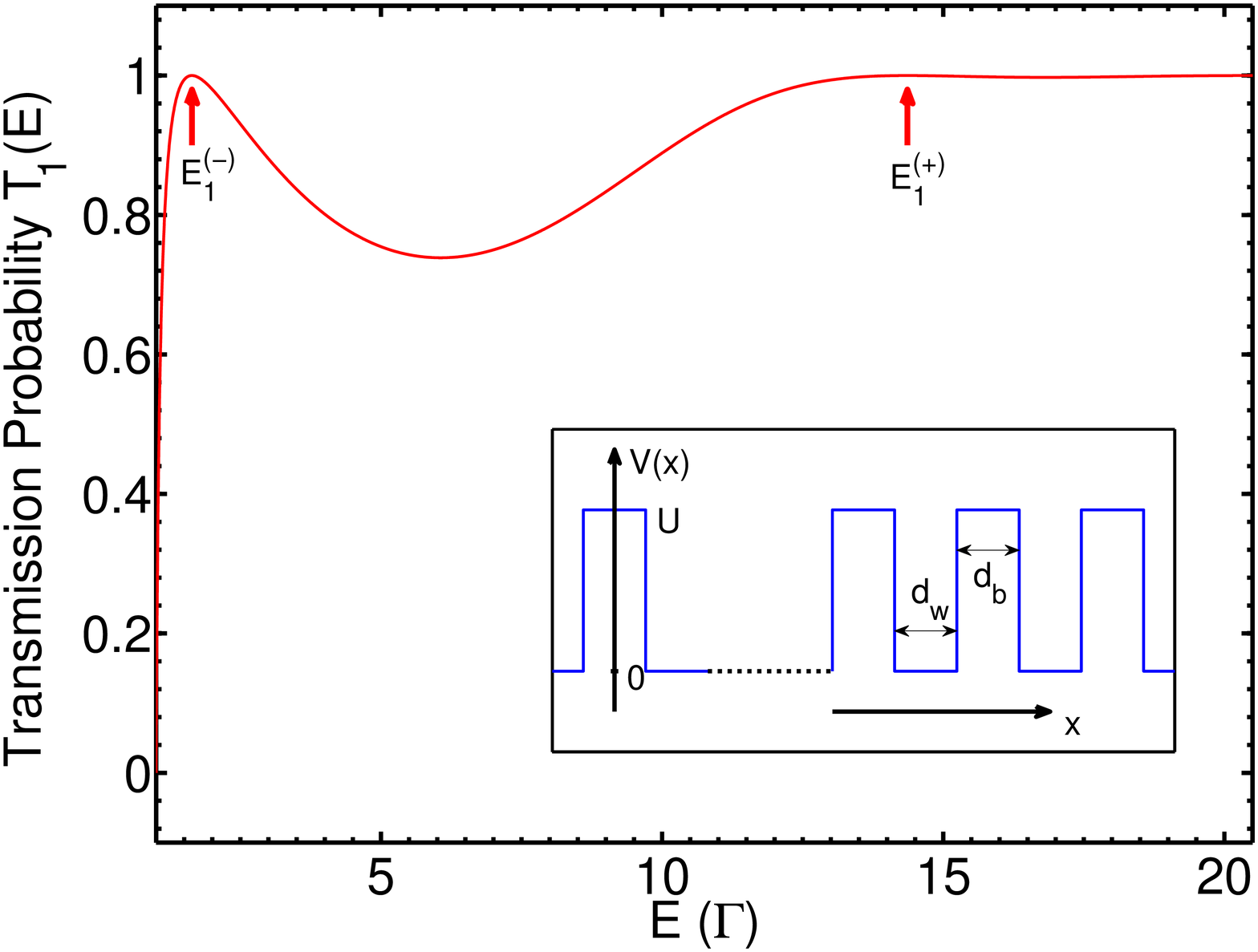}
\caption{
$(a)$ Transmission probability $T_1$ for a single rectangular barrier of $U =  8 \Gamma \equiv 8 (\hbar v_F / 2 d_B)$ and $d_B = 5 \ nm $ is plotted versus the incident energy $E$ [for reference: $\Gamma \approx 66 \ meV$ if $d_B = 5 \ nm$]; Arrows indicate the REs, $E_1^{(-)}$ and $E_1^{(+)}$, from eq.(\ref{eq4}); Inset: the rectangular potential model under study.
}
\label{fig1}
\end{center}
\end{figure}
\newpage

\begin{figure} [t]
\begin{center}
\includegraphics[width=12.0cm,height=9.0cm]{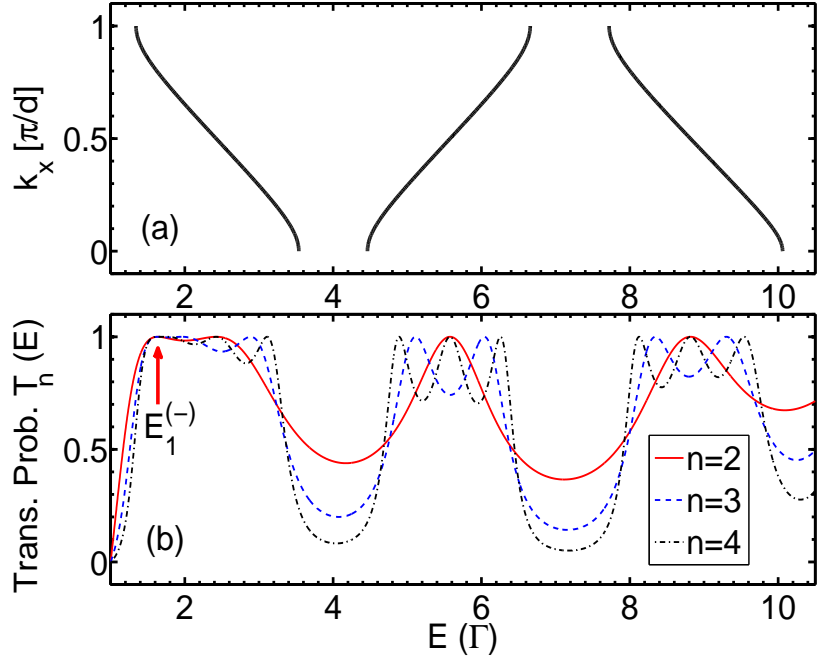}
\caption{
$(a)$ Cut of the band structure along the $(k_y = 0.1 nm^{-1})$-plane of the infinite EGSL with barrier parameters given in Fig.1 and $d_W = d_B$. $(b)$ Transmission probability $T_n$ of eq.(\ref{eq7}) is plotted as a function of the incident energy $E$ for finite EGSLs with different numbers of rectangular barriers $n$ [$U$ and $d_B = d_W$ are the same as in $(a)$]; Arrow indicates the barrier-induced RE $E_1^{(-)}$ which is completely insensitive to $n$ [Energy in units of $\Gamma \equiv \hbar v_F / 2 d_B$].
}
\label{fig2}
\end{center}
\end{figure}
\newpage

\begin{figure} [t]
\begin{center}
\includegraphics[width=12.0cm,height=9.0cm]{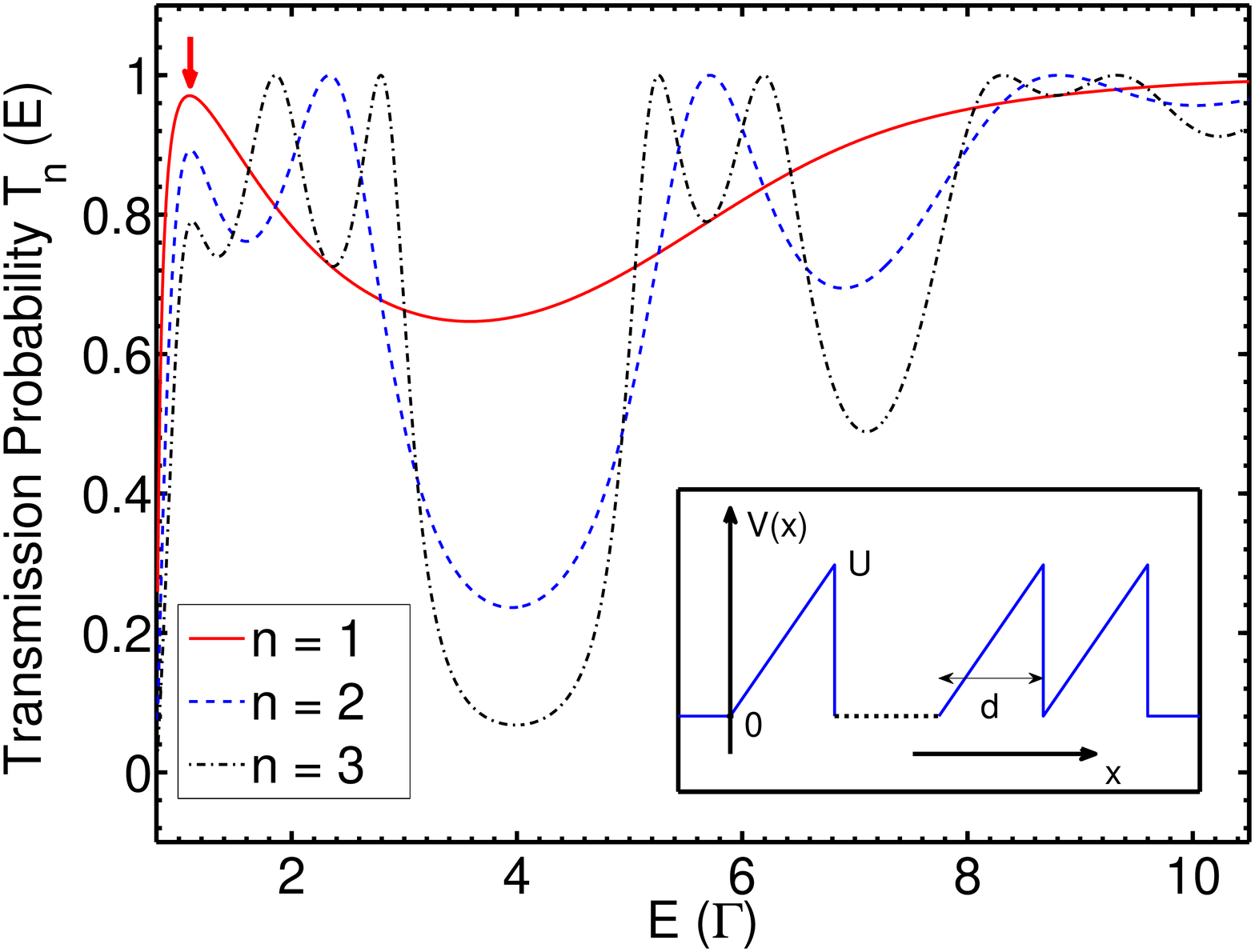}
\caption{
Transmission probability $T_n$ as a function of the incident energy $E$: numerical calculations for finite EGSLs with different numbers of triangular barriers $n$ [$U = 8 \Gamma \equiv 8 (\hbar v_F / d)$ and $d = 8 \ nm$]; Arrow indicates the barrier-induced RE which is insensitive to $n$ (Note: at this energy the transmission is imperfect and the resonance peaks become lower as $n$ increases). Inset: the triangular potential barrier model under study.
}
\label{fig3}
\end{center}
\end{figure}
\newpage

\begin{figure} [t]
\begin{center}
\includegraphics[width=12.0cm,height=9.0cm]{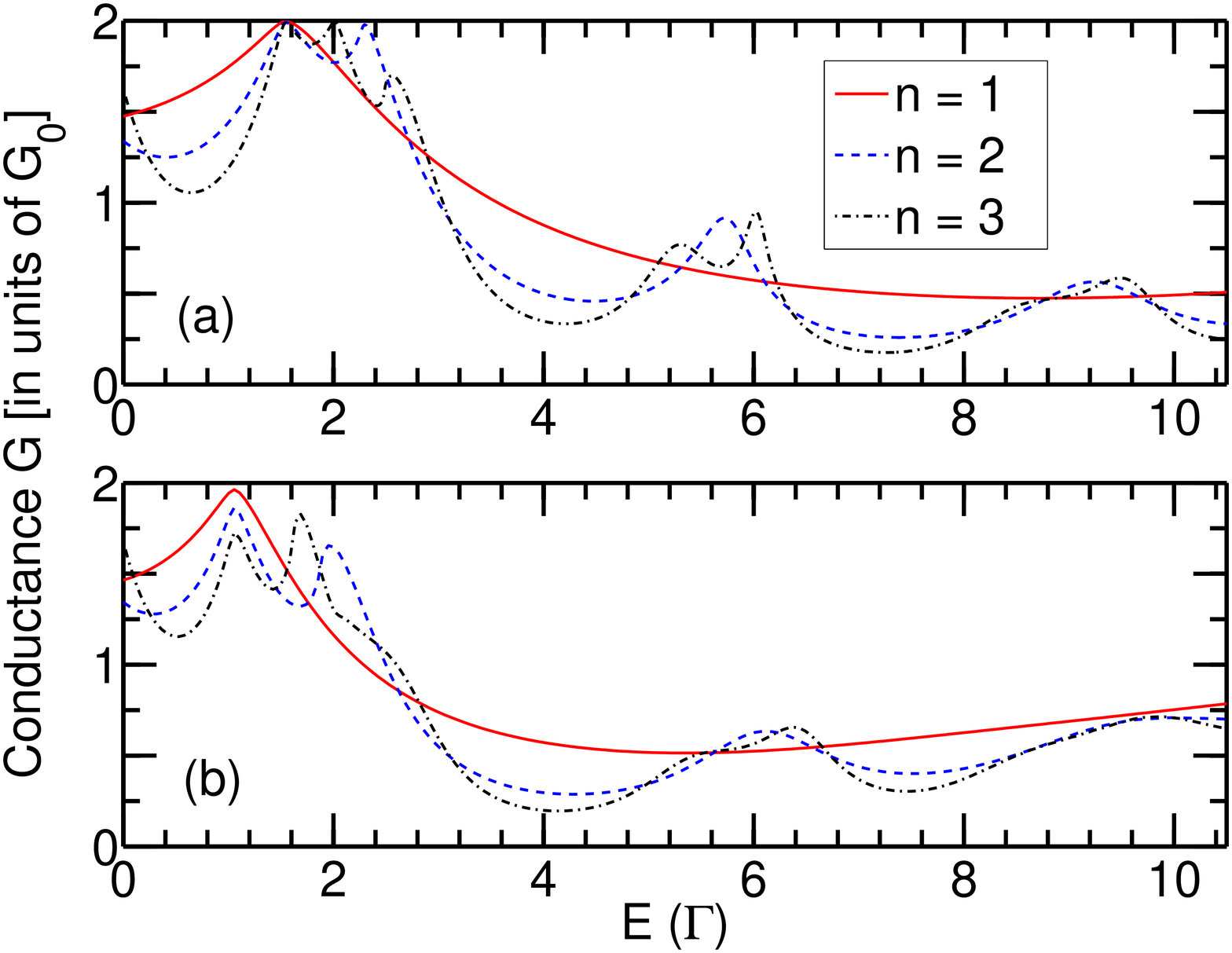}
\caption{
Conductance $G$ (in units of $G_0$) is plotted as a function of the incident energy $E$ (in units of $\Gamma$) for the finite EGSLs with different numbers of rectangular potential barriers $(a)$ [ $T_n$ given in Fig.2] or triangular potential barriers $(b)$ [$T_n$ given in Fig.3]. The $G(E)$-dependence adequately reflects the resonance behavior of $T_n (E)$.
}
\label{fig4}
\end{center}
\end{figure}
\newpage

\begin{figure} [t]
\begin{center}
\includegraphics[width=12.0cm,height=9.0cm]{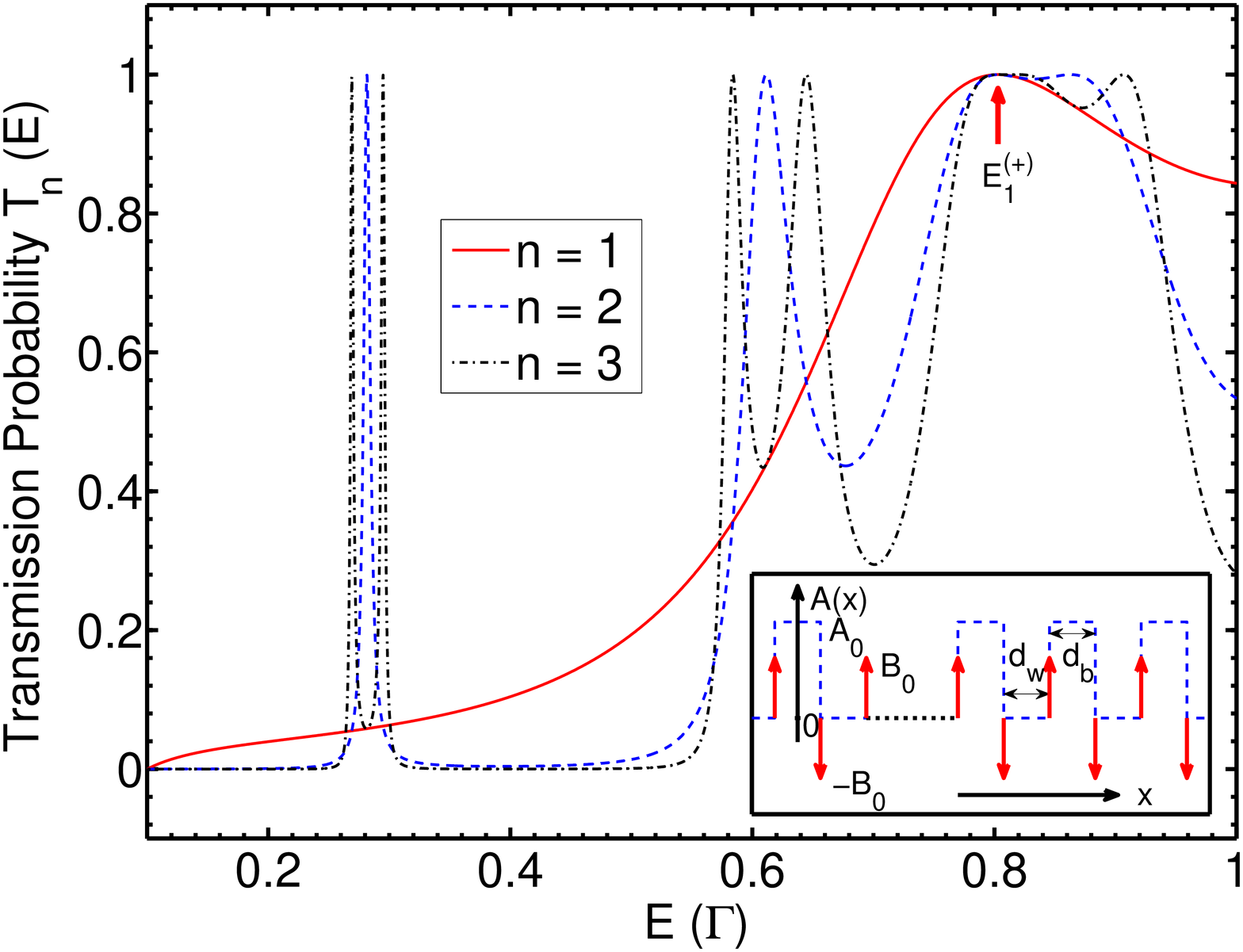}
\caption{
Transmission probability $T_n$ of eq.(\ref{eq19}) is plotted as a function of the incident energy $E$ for finite MGSLs with different numbers of $\delta$-function barriers $n$ [$A_0 = 0.4 $, $d_B = d_W = 5 \ nm$]; Arrow indicates the barrier-induced RE $E_1^{(+)}$ which is insensitive to $n$. Inset: the $\delta$-function magnetic potential barrier model under study.
}
\label{fig5}
\end{center}
\end{figure}
\newpage

\begin{figure} [t]
\begin{center}
\includegraphics[width=12.0cm,height=9.0cm]{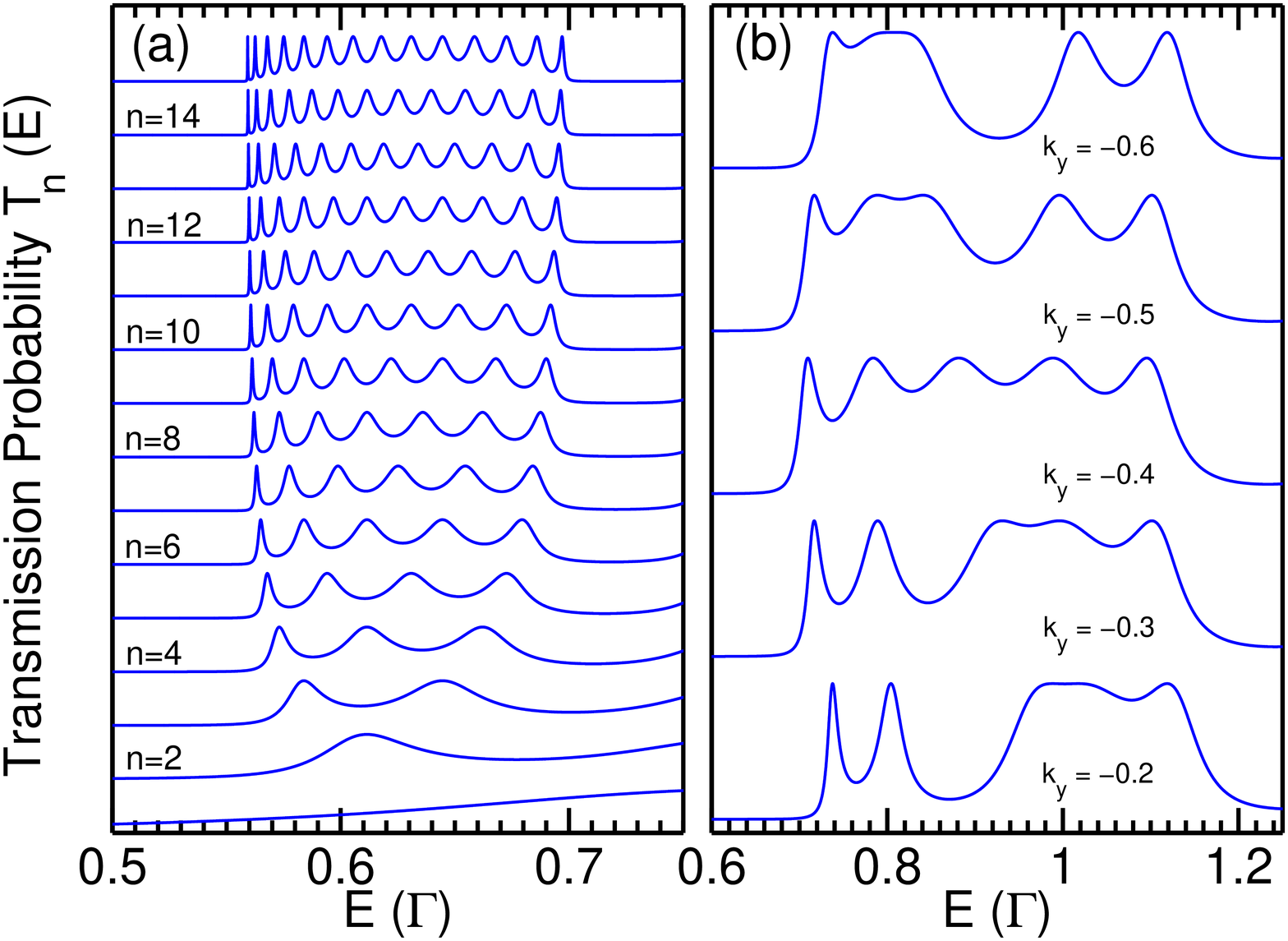}
\caption{
$(a)$ The $(n - 1)$-fold resonance splitting is in more detail demonstrated in a narrow energy range from Fig.5, but $n$ is now up to 15. $(b)$ To demonstrate the $k_y$-dependence of the resonance spectrum of transmission probability: $T_3 (E)$  for the same finite MGSL with $\delta$-function potential barriers [$A_0 = 0.8, \ d_B = d_W = 5 \ nm$, and $n = 3$] in the same energy range, but at different $k_y$-values in units of $nm^{-1}$ (given in the figure).
}
\label{fig6}
\end{center}
\end{figure}
\newpage

\end{document}